\documentclass[]{aa}
\usepackage{txfonts} 
\usepackage{graphicx}
\usepackage{natbib}

\begin{document}

   \title{Accretion and diffusion in white dwarfs}
    \subtitle{New diffusion timescales and applications to
          GD\,362 and G\,29-38}

   \author{D. Koester}
   \institute{   Institut f\"ur Theoretische Physik und Astrophysik, 
             University of Kiel, {D-24098 Kiel}, Germany
}

   \offprints{D. Koester\\ 
   \email{koester@astrophysik.uni-kiel.de}}

   \date{}

\authorrunning{D. Koester}

\titlerunning{Diffusion in white dwarfs}

\abstract{A number of cool white dwarfs with metal traces, of spectral
  types DAZ, DBZ, and DZ have been found to exhibit infrared excess
  radiation due to circumstellar dust. The origin of this dust is
  possibly a tidally disrupted asteroid that formed a debris disk now
  supplying the matter accreting onto the white dwarf. To reach any
  clear conclusions from the observed composition of the white dwarf
  atmosphere to that of the circumstellar matter, we need a detailed
  understanding of the accretion and diffusion process, in particular
  the diffusion timescales.}  {We aim to provide data for a wide range
  of white dwarf parameters and all possible observed chemical
  elements.}  {Starting from atmosphere models, we calculate the
  structure of the outer envelopes, obtaining the depth of the
  convection zone and the physical parameters at the lower
  boundary. These parameters are used to calculate the diffusion
  velocities using calculations of diffusion coefficients available in
  the literature.}  {With a simple example, we demonstrate that the
  observed element abundances are not identical to the accreted
  abundances. Reliable conclusions are possible only if we know or
  can assume that the star has reached a steady state between accretion
  and diffusion. In this case, most element abundances differ only by
  factors in the range 2-4 between atmospheric values and the circumstellar
  matter. Knowing the diffusion timescales, we can also
  accurately relate the accreted abundances to the observed
  ones. If accretion has stopped, or if the rates vary by large
  amounts, we cannot determine the composition of the accreted matter
  with any certainty.}{} \keywords{stars: white dwarfs -- stars:
  abundances -- accretion -- diffusion}

\maketitle

\section{Introduction}
Heavy elements in cool white dwarfs ($\la 25\,000$~K) should diffuse
downward in the atmospheres due to gravitational settling in the high
gravitational fields \citep{Schatzman45}. Nevertheless, one of the
first three ``classical'' white dwarfs, van Maanen~2, has been found
to exhibit strong \ion{Ca}{ii} resonance lines. This star has a
helium-rich atmosphere and is the prototype of the spectral class
DZ. Although an example of a star containing a hydrogen-rich
atmosphere with metal traces, G\,74-7, had been known for a
considerable period of time \citep[see the analysis by]
[]{Billeres.Wesemael.ea97}, it was not until 1997, that DAZ was
finally established as an important class
\citep{Koester.Provencal.ea97, Holberg.Barstow.ea97}, although DA are
the dominant spectral type of white dwarfs. The reason for the delay was
observational bias. Because of the much higher opacity of hydrogen at
low temperatures around 6000-20000\,K, metal spectral lines are
far weaker in a hydrogen atmosphere at the same
abundances. Large telescopes were therefore needed to detect DAZs in
significant numbers \citep{Zuckerman.Reid98, Zuckerman.Koester.ea03,
  Koester.Rollenhagen.ea05*b}.

Since diffusion timescales are always short compared to evolution
timescales, the metals must be supplied from a source from outside,
with the possible exception of carbon in the DQs, which can be
dredged-up from deeper layers \citep{Koester.Weidemann.ea82}. Assuming
the interstellar matter to be the outer source, this so-called
accretion/diffusion scenario was discussed in great detail in three
fundamental papers by \citet{Dupuis.Fontaine.ea92,
  Dupuis.Fontaine.ea93, Dupuis.Fontaine.ea93*b}. However, there are
some severe problems with this scenario: the apparent lack or at least
significant underabundance of hydrogen in the accreted matter, and the
lack of any correlation between the location of the white dwarfs and
the conditions of the ISM. Alternative scenarios were therefore
discussed, such as the accretion of comets \citep{Alcock.Fristrom.ea86}, or
a tidally disrupted asteroid \citep{Jura03}. A thorough
discussion of these alternative models was given by
\citet{Zuckerman.Koester.ea03}.

These objects have become the subject of renewed interest following
the detection of infrared excess radiation indicative of the presence
of circumstellar dust \citep{Zuckerman.Becklin87, Becklin.Farihi.ea05,
  Kilic.von-Hippel.ea05, Kilic.von-Hippel.ea06,
  von-Hippel.Kuchner.ea07, Jura.Farihi.ea07}. Although the spectra
from optical to IR can be fairly well modeled with simple disk models
\citep[e.g.,][]{Jura.Farihi.ea07}, the exact geometrical distribution
was only confirmed with the detection of gaseous metal disks around
hotter white dwarfs, which clearly show the signature of Keplerian
rotation \citep{Gansicke.Marsh.ea06}.

For all stars exhibiting metal traces in their atmospheres and
infrared excesses due to circumstellar dust, the source of the current
accretion is obviously the matter surrounding the star. The most
plausible origin -- and the only one discussed in the current
literature -- is an asteroid remaining from a former planetary system,
which has been tidally disrupted by the white dwarf. If metals are
found in the atmosphere, but no infrared excess, the situation is
unclear. However, \citet{Jura08} argues that in such a case the
circumstellar material may be provided by a higher number of small
asteroids instead of a single large one.

One should not forget, however, that the accretion process is not
well understood. We are not aware of a study of accretion from the ISM
for realistic conditions in the solar neighborhood, with a mixture of
gas and dust, and different phases of the ISM, but observationally we
know that on all scales in the universe, accretion is always
accompanied by the formation of an accretion disk. We should therefore
not exclude from our current considerations the possibility of some
kind of compromise between the current two models, accretion from the
ISM and accretion from the debris disk of a former asteroid.

It is widely accepted that accretion from the ISM occurs for almost
all white dwarfs during their lifetimes because more than half of all
DB (and with higher quality observations perhaps all) show traces of
hydrogen, which according to our current understanding can only be
accreted \citep{Voss.Koester.ea07}. This is also true for the
significant amount of hydrogen detected in GD\,362. If the accreted
metals indeed originate in a disrupted asteroid, one has to assume an
independent source for the hydrogen, unless the asteroid contained a
substantial amount of water \citep{Jura.Muno.ea09}. The approximately
solar ratio of metals to hydrogen (see below) is in both scenarios
purely accidental.

An important tool in understanding the origin of this dust is the
determination of its composition by the analysis of the atmospheric
abundances. The latter is straightforward (if high-resolution, high
signal-to-noise spectra can be obtained!), but the connection between
atmospheric and accreted composition requires a detailed understanding
of the accretion and diffusion processes. The composition of the
accreting matter need not necessarily be identical to that of the
circumstellar matter \citep[see e.g.,][]{Alcock.Illarionov80}. And the
composition observed in the stellar atmosphere is modified by
diffusion out of the outer reservoir, which occurs at different rates
(diffusion timescales) for different elements. This is the part of the
problem studied in this paper, which extends the work of
\citet{Koester.Wilken06}.

Since the study of \citet{Koester.Wilken06}, infrared excess radiation
has been identified in many more hydrogen-rich and helium-rich white
dwarfs with metals. Improved observations have allowed us to identify 15
heavy elements and place upper limits on a few more in GD\,362
\citep{Zuckerman.Koester.ea07}. For many of those elements, diffusion
timescales are unavailable in the literature, inhibiting any firm
conclusions about the composition of the circumstellar matter. We
therefore calculated atmosphere and envelope models for hydrogen and
helium atmosphere white dwarfs throughout the interesting temperature
range and calculated the timescales for many elements. In this work,
we follow the ground work laid by the Montreal group, using their data
on the Coulomb collision integrals \citep{Paquette.Pelletier.ea86} and
to a large extent the methods outlined in
\citet{Paquette.Pelletier.ea86*b}.

\section{Envelope models}
To determine diffusion timescales, we need to model the structure of
the outer layers of the white dwarfs, including the complete
convection zone.  The outer boundary is determined from atmospheric
models, which assume  Local Thermodynamic Equilibrium
(LTE), hydrostatic equilibrium, convective energy transport, and a
detailed calculation of the radiative energy transfer. An up-to-date
description of input physics, data,  and numerical methods is given by
\cite{Koester09}. The parameters of these models are effective
temperature $T_\mathrm{eff}$\ and surface gravity $\log g$, in addition to the
element abundances. The models are calculated down into the star to very large
Rosseland optical depth $\tau_\mathrm{R}$, usually between 1000 and 1500.

The value of physical parameters such as pressure, temperature, and density,
at some defined value $\tau_\mathrm{R}$, which is a free
parameter chosen to be between 1 and 500, is the starting point for the
integration of the deeper layers. In this integration, we also need to know
the mass and radius of the star, which were obtained from the finite
temperature mass-radius relations of \citet{Wood95}.

We use the standard equations of stellar structure for the
conservation of mass, hydrostatic equilibrium, and energy transfer. The
fourth equation, of energy conservation, is replaced by the assumption
that the energy flux $l(r)$ at the radius $r$ is proportional to the
mass $m(r)$ inside that radius
\begin{equation}
l(r) = \frac{L}{M}\,m(r) ,
\end{equation}
for total mass $M$ and luminosity $L$. This is almost equivalent
to assuming a constant luminosity throughout the outer layers. The
equations are integrated inward from the boundary condition until the
fractional mass in the envelope $\Delta M /M = (1 - m(r)/M) $ reaches
approximately $10^{-4}$. The algorithms and variable definitions, as
well as the program code, are taken in large part from our white
dwarf structure code, which has been used in numerous
projects. Updated input physics included in our code, are the following:
\begin{itemize}
\item the equation of state (EOS) is that of \citet{Saumon.Chabrier.ea95},
  which is adapted from the tables obtained from the AAS CD-ROM series Vol. 5
  (1995). This is probably the most sophisticated EOS for
  hydrogen/helium mixtures available today.
\item absorption coefficients were obtained from the OPAL project
\citep{Iglesias.Rogers96}, supplemented at low temperatures with the
tables of \citet{Ferguson.Alexander.ea05}. 
\item thermal conductivity data are from \cite{Potekhin.Baiko.ea99}; the 
tables were obtained from their website.\footnote{http://www.ioffe.rssi.ru/astro/conduct}
\end{itemize}

The OPAL tables were slightly extended at high densities by data
calculated for H/He mixtures with the help of the Los Alamos TOPS
program on their web-site.\footnote{
    http://www.t4.lanl.gov/cgi-bin/opacity/astro.pl} Nevertheless,
in some of our envelope models, the structure falls into a regime of
the opacity tables, where the assumptions become invalid and the data
are unreliable. In this case, the highest reliable density value at
the given temperature was used in the extrapolation. Fortunately this
had no effect, since in this regime the opacity is always high and the
models are convective, the temperature gradient being practically
adiabatic. We confirmed this finding by using instead old Los Alamos
opacity tables \citep{Cox.Stewart70}, and no significant changes to
the models were found.

\subsection{The convection zone}
The convection zone in white dwarfs -- both the atmosphere and
envelope -- is calculated with a version of the mixing-length
approximation. It has become customary to denote the specific version
as, e.g., ML2/$\alpha=0.6$ or in shorthand as ML2/0.6. Here ML2
represents a choice of three dimensionless parameters, which are
defined and explained in \citet{Fontaine.Villeneuve.ea81} and
\citet{Tassoul.Fontaine.ea90}, the number (0.6) is the mixing-length
as a multiple of the pressure scale height. \cite{Koester.Allard.ea94}
and \cite{Bergeron.Wesemael.ea95} demonstrated that a version with
``intermediate'' efficiency of energy transport is the best choice for
DAs in ensuring a consistent fit to optical and UV spectra
simultaneously. The Bergeron choice ML2/0.6 is the de-facto standard
for DA model atmospheres today, and our ``standard'' grid in this
paper also uses this version in atmosphere and envelope calculations.

To our knowledge there is no comparable published study of DB white
dwarfs. However, there is in \citet{Beauchamp.Wesemael.ea99} a reference to
spectroscopic fits of optical and UV data, which favors the version ML2/1.25.
We note that these conclusions for DAs and DBs are based
on observed spectra and thus describe the atmospheric layers, which
produce the emerging light, i.e., above $ \tau_\mathrm{R} \approx 1$.

It is well established that the deeper layers of the atmosphere and
envelope are not necessarily correctly described by the same version
of the mixing-length approximation.  The atmospheric parameters of
G\,29-38 are $T_\mathrm{eff}$\ = 11\,700~K, $\log g$\ = 8.10, when
analyzed with ML2/0.6 models. The convection zone in such a model is
extremely thin, and has  a lower boundary at around optical depth 10. Such a
thin cvz has a thermal timescale, defined as
\begin{equation}
     \tau_\mathrm{th} =  \frac{\int_0^{M_\mathrm{cvz}} c_P T dm}{L}
\end{equation}
of less than 1s. However, G\,29-38 is a variable ZZ Ceti star, with a
shortest period a little larger than 200s. \cite{Winget.van-Horn.ea83}
and \citet{Tassoul.Fontaine.ea90} argued that the thermal
timescale of the cvz should be similar to the period, that is the cvz
should be much deeper than indicated by the atmospheric ML
parameters. The same conclusion is reached by fitting the non-linear
light-curve; the thermal timescale at 11\,700~K is predicted to be
$\approx 100$s \citep{Montgomery05}. 

From a completely different point of view, a similar conclusion was
reached by \cite{Ludwig.Jordan.ea94}. By comparing the mean
temperature structure in two-dimensional hydrodynamic simulations of
the outer layers of a DA white dwarf with model atmospheres using MLT,
they found that no single MLT version can describe the entire structure of
the cvz. The efficiency, or in the MLT parameterization the
mixing-length, must increase with depth.

We therefore calculated a second set of DA envelopes, where the MLT
version was switched to ML2/2.0 in the envelope calculation. The
result depends strongly on the exact structure of the atmosphere model
and the location of the ``matching point'', the starting point in the
envelope integration. If this layer is below the atmospheric cvz,
e.g., deeper than $\tau_\mathrm{R} \approx 10$ in the case of the
G29-38 model, or at such large depth that the convection is nearly
adiabatic, the envelope integration will not produce any deepening of
the cvz, regardless of the efficiency of the MLT version. The matching
must occur within the super-adiabatic part of the atmospheric
convection zone for the change in efficiency to influence the model
structure. We considered the optical depth of this layer as a free
parameter and calibrated it by demanding a thermal timescale of
$\approx 100$s for the G29-38 model. This is of course unsatisfactory,
and in the future we hope to find a more consistent description with a
smooth change of efficiency from shallow to deeper
layers. Nevertheless, our choice is supported by the pulsation
properties of the DAs and should provide more realistic estimates of
diffusion timescales.

The situation is more favorable for the DB stars than for the DAs.
\cite{Benvenuto.Althaus97} compared results for thermal timescales and
cvz depths in DBs between the more sophisticated convection theory of
\citet{Canuto.Mazzitelli91,Canuto.Mazzitelli92} and \citet{Canuto.Goldman.ea96},
with different simple MLT versions. They concluded that a convective
efficiency between ML2/1.0 and ML2/2.0 reproduces the results and
predicts the correct location of the blue edge of the DB instability
strip. This was confirmed by \cite{Corsico.Althaus.ea08}, who
concluded that only ML2/1.25 predicts the correct location. A
similar result was also obtained from the light-curve fitting of the
prototype variable DB star GD\,358 \citep{Montgomery07}.
We thus chose this version for both the atmosphere and  the envelope
calculations.

More fundamentally, it is well known that the MLT
approximation provides a poor description of the various aspects of a
real convection zone. For the case of DA white dwarfs, this was
studied with an extensive comparison of two-dimensional
radiation-hydrodynamic simulations with 1D structures by
\cite{Freytag.Ludwig.ea96}. Even the definition of the lower boundary
is ambiguous: depending on whether one uses the classical stability
criterion, the layers with significant convective flux, or the layers
with non-zero velocities, the resulting mass in the ``convection
zone'' can differ by orders of magnitude. While the temperature
structure (related to convective flux) is probably the most important
quantity for the pulsational properties, for diffusion timescales,
the mixed region with non-zero velocity is relevant. In the example
studied by \citet{Freytag.Ludwig.ea96}, the mass in the latter is 300
times higher than in the former. 

The general result that the velocity field extends far below the lower
limit of the unstable region and even the flux overshoot was also
confirmed by \citet{Montgomery.Kupka04} with their non-local model of
convection in DAs.  This implies that the
diffusion timescales in stars with convection zones may be orders of
magnitude larger than estimated with our current MLT approximations.

\begin{table}[ht]
\caption{Conditions at the bottom of the convection zones in DA models
  with surface gravity $\log g = 8$ and the standard mixing length
  version ML2/0.6 for atmosphere and envelope. The second column is
  the fractional mass $q = \log M_\mathrm{cvz}/M$ in
  the zone or down to an optical depth $\tau_R = 5$, whichever is deeper.
  The next columns give the logarithms of gas pressure, temperature,
  and mass density at this level. All physical quantities are in cgs
  units. \label{cvzdepthsa}}
\begin{center}
\begin{tabular}{rrrrr}
  \hline
  \noalign{\smallskip}
  $T_\mathrm{eff}$  &  $\log M_\mathrm{cvz}/M$ & $\log P $ & $\log T$  & $\log
  \rho$   \\
  \noalign{\smallskip}
  \hline
   6000. &   -7.570 &  14.510 &   5.914 &  0.393\\
   7000. &   -8.423 &  13.655 &   5.758 & -0.293\\
   8000. &   -9.002 &  13.076 &   5.665 & -0.773\\
   9000. &   -9.719 &  12.359 &   5.543 & -1.371\\
   9500. &  -10.278 &  11.799 &   5.438 & -1.826\\
  10000. &  -11.063 &  11.014 &   5.289 & -2.466\\
  10500. &  -12.094 &   9.983 &   5.098 & -3.310\\
  11000. &  -13.392 &   8.684 &   4.870 & -4.386\\
  11100. &  -13.681 &   8.395 &   4.821 & -4.629\\
  11200. &  -13.975 &   8.101 &   4.771 & -4.876\\
  11300. &  -14.288 &   7.788 &   4.718 & -5.136\\
  11400. &  -14.638 &   7.438 &   4.653 & -5.422\\
  11500. &  -15.049 &   7.026 &   4.559 & -5.739\\
  11600. &  -15.427 &   6.649 &   4.456 & -6.013\\
  11700. &  -15.675 &   6.401 &   4.393 & -6.196\\
  11800. &  -15.850 &   6.225 &   4.353 & -6.330\\
  11900. &  -15.947 &   6.129 &   4.332 & -6.404\\
  12000. &  -16.006 &   6.070 &   4.321 & -6.450\\
  12500. &  -16.065 &   6.010 &   4.316 & -6.504\\
  13000. &  -16.070 &   6.006 &   4.326 & -6.521\\
  14000. &  -16.009 &   6.067 &   4.361 & -6.500\\
  15000. &  -15.949 &   6.126 &   4.391 & -6.472\\
  16000. &  -15.891 &   6.185 &   4.420 & -6.444\\
  17000. &  -15.839 &   6.237 &   4.447 & -6.420\\
  18000. &  -15.794 &   6.282 &   4.473 & -6.400\\
  19000. &  -15.754 &   6.322 &   4.496 & -6.385\\
  20000. &  -15.719 &   6.357 &   4.518 & -6.371\\
  21000. &  -15.687 &   6.389 &   4.538 & -6.361\\
  22000. &  -15.658 &   6.417 &   4.558 & -6.351\\
  23000. &  -15.632 &   6.444 &   4.575 & -6.343\\
  24000. &  -15.607 &   6.469 &   4.592 & -6.335\\
  25000. &  -15.584 &   6.492 &   4.609 & -6.328\\
\noalign{\smallskip}
\hline
\end{tabular}
\end{center}
\end{table}

\begin{table}[ht]
\caption{Similar to Table~\ref{cvzdepthsa}, but with a more efficient
  convective energy transport in the deeper layers, as explained in
  the text. For $T_\mathrm{eff}$ $>$ 13000\,K the entries are identical with
  Table~\ref{cvzdepthsa} and are not repeated here. \label{cvzdepthsb}}
\begin{center}
\begin{tabular}{rrrrr}
  \hline
  \noalign{\smallskip}
  $T_\mathrm{eff}$  &  $\log M_\mathrm{cvz}/M$ & $\log P $ & $\log T$  & $\log
  \rho$   \\
  \noalign{\smallskip}
  \hline
   6000. &   -7.564 &  14.516 &   5.916 &  0.398\\
   7000. &   -8.376 &  13.703 &   5.769 & -0.256\\
   8000. &   -8.876 &  13.202 &   5.691 & -0.673\\
   9000. &   -9.429 &  12.649 &   5.602 & -1.140\\
   9500. &   -9.803 &  12.275 &   5.536 & -1.448\\
  10000. &  -10.328 &  11.749 &   5.437 & -1.877\\
  10500. &  -11.040 &  11.037 &   5.303 & -2.457\\
  11000. &  -11.945 &  10.132 &   5.135 & -3.198\\
  11100. &  -12.151 &   9.926 &   5.097 & -3.367\\
  11200. &  -12.364 &   9.712 &   5.058 & -3.543\\
  11300. &  -12.587 &   9.490 &   5.019 & -3.727\\
  11400. &  -12.819 &   9.258 &   4.978 & -3.919\\
  11500. &  -13.058 &   9.018 &   4.936 & -4.118\\
  11600. &  -13.301 &   8.775 &   4.895 & -4.320\\
  11700. &  -13.544 &   8.532 &   4.853 & -4.522\\
  11800. &  -13.806 &   8.270 &   4.809 & -4.743\\
  11900. &  -13.996 &   8.080 &   4.777 & -4.902\\
  12000. &  -14.229 &   7.847 &   4.738 & -5.097\\
  12500. &  -15.555 &   6.521 &   4.437 & -6.122\\
  13000. &  -16.070 &   6.006 &   4.326 & -6.521\\
\noalign{\smallskip}
\hline
\end{tabular}
\end{center}
\end{table}

\begin{table}[ht]
\caption{Similar to Table~\ref{cvzdepthsa}, but for DB white dwarfs. 
Convection parameters are ML2/1.25 for atmosphere and envelope models.
\label{cvzdepthsc}}
\begin{center}
\begin{tabular}{rrrrr}
  \hline
  \noalign{\smallskip}
  $T_\mathrm{eff}$  &  $\log M_\mathrm{cvz}/M$ & $\log P $ & $\log T$  & $\log
  \rho$   \\
  \noalign{\smallskip}
  \hline
   6000. &   -4.840 &  17.243 &   5.938 &  2.868\\
   7000. &   -4.754 &  17.330 &   6.098 &  2.909\\
   8000. &   -4.760 &  17.324 &   6.275 &  2.882\\
   9000. &   -4.789 &  17.296 &   6.440 &  2.825\\
  10000. &   -4.833 &  17.253 &   6.551 &  2.754\\
  11000. &   -4.954 &  17.131 &   6.598 &  2.636\\
  12000. &   -5.132 &  16.954 &   6.603 &  2.486\\
  13000. &   -5.332 &  16.752 &   6.589 &  2.322\\
  14000. &   -5.545 &  16.538 &   6.565 &  2.151\\
  15000. &   -5.780 &  16.303 &   6.532 &  1.964\\
  16000. &   -6.059 &  16.023 &   6.486 &  1.743\\
  17000. &   -6.413 &  15.668 &   6.421 &  1.462\\
  18000. &   -6.884 &  15.196 &   6.331 &  1.086\\
  19000. &   -7.572 &  14.507 &   6.202 &  0.531\\
  20000. &   -8.657 &  13.421 &   5.992 & -0.333\\
  21000. &  -10.046 &  12.031 &   5.727 & -1.470\\
  22000. &  -11.368 &  10.708 &   5.493 & -2.567\\
  23000. &  -11.581 &  10.495 &   5.463 & -2.751\\
  24000. &  -11.746 &  10.330 &   5.441 & -2.895\\
  25000. &  -11.911 &  10.165 &   5.419 & -3.038\\
\noalign{\smallskip}
\hline
\end{tabular}
\end{center}
\end{table}

\section{Diffusion coefficients and timescales}
For the calculation of diffusion timescales we follow closely
the fundamental works of \citet{Paquette.Pelletier.ea86} and
\citet{Paquette.Pelletier.ea86*b}. From the tables in the first paper,
we take the fit coefficients for the calculation of Coulomb collision
integrals and the diffusion coefficients as well as thermal diffusion
coefficients. The equations to calculate diffusion velocities are
taken from the second paper with two minor modifications. First, we
use what the authors call the ``second method'' for the thermal
diffusion coefficient. This is based on the approach by
\citet{Chapman.Cowling70}, and the necessary data can be found in the
first paper cited above. The second change is purely cosmetic. 
To determine the effective charge of the trace element 2, we
calculate an effective ionization potential $\chi_\mathrm{eff}$ as
described in their Eqs. 19-21, and compare this with the true
ionization potentials $\chi(Z)$ of the ions of element 2 with charge
$Z$.  Rather than taking the effective charge as $Z$, if
$\chi_\mathrm{eff}(Z)/\chi(Z) \le 1$ {\em and}
$\chi_\mathrm{eff}(Z+1)/\chi(Z+1)> 1 $, we interpolate a non-integer
$Z_\mathrm{eff}$ between $Z$ and $Z+1$. Diffusion coefficients and
diffusion velocities are then calculated for ions $Z$ and $Z+1$ and a
weighted average taken depending on the value of $Z_\mathrm{eff}$. The
only purpose of this approach is to avoid unphysical wiggles and steps in
the relation between diffusion timescales and effective temperatures (see
e.g., Fig.~4 in \citet{Paquette.Pelletier.ea86*b}).

The diffusion velocity $v_\mathrm{diff}$ is obtained using Eq.~4 in
\citet{Paquette.Pelletier.ea86*b}, neglecting the concentration term,
since we consider only diffusion of trace elements. The diffusion
time scale is
\begin{equation}
      \tau_\mathrm{diff} = \frac{M_\mathrm{cvz}}{4\pi r^2\, \rho\, 
v_\mathrm{diff}}
\end{equation}
where $r$ is the local radius at the bottom of the convection zone and
$\rho$ is the local mass density. Results for the stellar
models in Tables~\ref{cvzdepthsa} to \ref{cvzdepthsc} are given in
Tables~\ref{taua} to \ref{tauc} for six important elements; data for
other models and/or other elements can be requested from the author.

\begin{table}[ht]
\caption{Diffusion timescales at the base of the convection zone (or
  at $\tau_\mathrm{R} = 5$) for DA models with $\log g$\ = 8, using
  the standard assumption of ML2/0.6 for atmosphere and envelope
  models. Columns 2-6 give the logarithm of the timescales in years
  for the elements C, Na, Mg, Si, Ca, and Fe.
\label{taua}}
\begin{center}
\begin{tabular}{rrrrrrr}
  \hline
  \noalign{\smallskip}
  $T_\mathrm{eff}$  & C & Na & Mg & Si & Ca & Fe\\
  \noalign{\smallskip}
  \hline
   6000.0 &  4.50 &   4.34 &   4.29 &   4.21 &   4.14 &   4.00\\
   7000.0 &  3.88 &   3.60 &   3.54 &   3.52 &   3.45 &   3.33\\
   8000.0 &  3.49 &   3.16 &   3.07 &   3.12 &   3.06 &   2.95\\
   9000.0 &  3.03 &   2.70 &   2.60 &   2.66 &   2.61 &   2.50\\
   9500.0 &  2.68 &   2.31 &   2.21 &   2.30 &   2.24 &   2.13\\
  10000.0 &  2.18 &   1.74 &   1.65 &   1.79 &   1.68 &   1.59\\
  10500.0 &  1.55 &   0.98 &   0.95 &   1.15 &   0.94 &   0.88\\
  11000.0 &  0.55 &  -0.00 &   0.05 &   0.32 &  -0.04 &  -0.03\\
  11100.0 &  0.33 &  -0.22 &  -0.13 &   0.14 &  -0.25 &  -0.24\\
  11200.0 &  0.10 &  -0.46 &  -0.33 &  -0.07 &  -0.46 &  -0.44\\
  11300.0 & -0.14 &  -0.72 &  -0.55 &  -0.33 &  -0.69 &  -0.67\\
  11400.0 & -0.41 &  -1.01 &  -0.79 &  -0.61 &  -0.94 &  -0.92\\
  11500.0 & -0.74 &  -1.35 &  -1.07 &  -0.98 &  -1.25 &  -1.22\\
  11600.0 & -1.03 &  -1.64 &  -1.32 &  -1.31 &  -1.51 &  -1.57\\
  11700.0 & -1.35 &  -1.84 &  -1.50 &  -1.52 &  -1.70 &  -1.80\\
  11800.0 & -1.54 &  -1.97 &  -1.63 &  -1.66 &  -1.83 &  -1.95\\
  11900.0 & -1.65 &  -2.05 &  -1.70 &  -1.74 &  -1.90 &  -2.04\\
  12000.0 & -1.70 &  -2.10 &  -1.74 &  -1.79 &  -1.95 &  -2.09\\
  12500.0 & -1.74 &  -2.14 &  -1.78 &  -1.83 &  -1.99 &  -2.12\\
  13000.0 & -1.73 &  -2.14 &  -1.79 &  -1.83 &  -1.99 &  -2.12\\
  14000.0 & -1.61 &  -2.08 &  -1.73 &  -1.76 &  -1.93 &  -2.04\\
  15000.0 & -1.48 &  -2.03 &  -1.69 &  -1.70 &  -1.89 &  -1.97\\
  16000.0 & -1.36 &  -1.98 &  -1.65 &  -1.64 &  -1.84 &  -1.90\\
  17000.0 & -1.32 &  -1.94 &  -1.62 &  -1.58 &  -1.80 &  -1.83\\
  18000.0 & -1.28 &  -1.90 &  -1.59 &  -1.53 &  -1.77 &  -1.77\\
  19000.0 & -1.24 &  -1.86 &  -1.57 &  -1.48 &  -1.74 &  -1.71\\
  20000.0 & -1.21 &  -1.83 &  -1.55 &  -1.43 &  -1.72 &  -1.68\\
  21000.0 & -1.18 &  -1.80 &  -1.53 &  -1.38 &  -1.69 &  -1.66\\
  22000.0 & -1.15 &  -1.77 &  -1.51 &  -1.34 &  -1.67 &  -1.64\\
  23000.0 & -1.12 &  -1.74 &  -1.50 &  -1.31 &  -1.65 &  -1.61\\
  24000.0 & -1.09 &  -1.70 &  -1.48 &  -1.27 &  -1.63 &  -1.59\\
  25000.0 & -1.06 &  -1.67 &  -1.47 &  -1.24 &  -1.60 &  -1.57\\
\noalign{\smallskip}
\hline
\end{tabular}
\end{center}
\end{table}

\begin{table}[ht]
\caption{Similar to Table~\ref{taua}, but with more efficient
  convection (ML2/2.0) below $\tau_\mathrm{R} = 2$, leading to deeper
convection zones in closer agreement with expectations for the variable
ZZ Cetis. 
\label{taub}}

\begin{center}
\begin{tabular}{rrrrrrr}
  \hline
  \noalign{\smallskip}
  $T_\mathrm{eff}$  & C & Na & Mg & Si & Ca & Fe\\
  \noalign{\smallskip}
  \hline
   6000.0 &  4.50 &   4.34 &   4.29 &   4.21 &   4.15 &   4.01\\
   7000.0 &  3.91 &   3.64 &   3.58 &   3.55 &   3.49 &   3.37\\
   8000.0 &  3.57 &   3.25 &   3.16 &   3.20 &   3.14 &   3.03\\
   9000.0 &  3.22 &   2.91 &   2.81 &   2.85 &   2.82 &   2.70\\
   9500.0 &  2.98 &   2.67 &   2.56 &   2.61 &   2.57 &   2.46\\
  10000.0 &  2.65 &   2.30 &   2.20 &   2.27 &   2.22 &   2.11\\
  10500.0 &  2.19 &   1.78 &   1.68 &   1.81 &   1.72 &   1.62\\
  11000.0 &  1.64 &   1.12 &   1.05 &   1.24 &   1.07 &   0.99\\
  11100.0 &  1.51 &   0.95 &   0.91 &   1.11 &   0.92 &   0.85\\
  11200.0 &  1.37 &   0.79 &   0.77 &   0.98 &   0.75 &   0.70\\
  11300.0 &  1.19 &   0.61 &   0.61 &   0.84 &   0.58 &   0.54\\
  11400.0 &  1.00 &   0.43 &   0.45 &   0.69 &   0.39 &   0.38\\
  11500.0 &  0.82 &   0.25 &   0.29 &   0.54 &   0.22 &   0.22\\
  11600.0 &  0.64 &   0.07 &   0.12 &   0.38 &   0.03 &   0.04\\
  11700.0 &  0.45 &  -0.10 &  -0.03 &   0.23 &  -0.14 &  -0.13\\
  11800.0 &  0.25 &  -0.30 &  -0.21 &   0.06 &  -0.33 &  -0.31\\
  11900.0 &  0.10 &  -0.46 &  -0.34 &  -0.07 &  -0.47 &  -0.45\\
  12000.0 & -0.08 &  -0.66 &  -0.50 &  -0.26 &  -0.64 &  -0.62\\
  12500.0 & -1.15 &  -1.74 &  -1.42 &  -1.41 &  -1.61 &  -1.68\\
  13000.0 & -1.73 &  -2.14 &  -1.79 &  -1.83 &  -1.99 &  -2.12\\
\noalign{\smallskip}
\hline
\end{tabular}
\end{center}
\end{table}

\begin{table}[ht]
\caption{Similar to Table~\ref{taua}, but for non-DA models without
  outer hydrogen layer (DB, DC).
\label{tauc}}
\begin{center}
\begin{tabular}{rrrrrrr}
  \hline
  \noalign{\smallskip}
  $T_\mathrm{eff}$  & C & Na & Mg & Si & Ca & Fe\\
  \noalign{\smallskip}
  \hline
   6000.0 &  6.50 &   6.57 &   6.55 &   6.59 &   6.56 &   6.56\\
   7000.0 &  6.48 &   6.52 &   6.54 &   6.52 &   6.53 &   6.49\\
   8000.0 &  6.40 &   6.40 &   6.41 &   6.40 &   6.41 &   6.33\\
   9000.0 &  6.34 &   6.32 &   6.32 &   6.32 &   6.32 &   6.22\\
  10000.0 &  6.31 &   6.28 &   6.28 &   6.29 &   6.27 &   6.15\\
  11000.0 &  6.25 &   6.21 &   6.22 &   6.20 &   6.18 &   6.06\\
  12000.0 &  6.17 &   6.13 &   6.13 &   6.09 &   6.05 &   5.94\\
  13000.0 &  6.06 &   6.01 &   5.98 &   5.96 &   5.88 &   5.80\\
  14000.0 &  5.96 &   5.88 &   5.85 &   5.85 &   5.72 &   5.64\\
  15000.0 &  5.84 &   5.72 &   5.71 &   5.71 &   5.53 &   5.46\\
  16000.0 &  5.68 &   5.54 &   5.54 &   5.49 &   5.33 &   5.24\\
  17000.0 &  5.51 &   5.35 &   5.32 &   5.23 &   5.11 &   4.98\\
  18000.0 &  5.26 &   5.05 &   4.99 &   4.89 &   4.83 &   4.64\\
  19000.0 &  4.73 &   4.62 &   4.56 &   4.44 &   4.43 &   4.22\\
  20000.0 &  4.06 &   3.99 &   3.91 &   3.77 &   3.78 &   3.56\\
  21000.0 &  3.22 &   3.12 &   3.00 &   2.86 &   2.96 &   2.68\\
  22000.0 &  2.36 &   2.15 &   2.05 &   1.89 &   2.03 &   1.82\\
  23000.0 &  2.21 &   2.00 &   1.90 &   1.74 &   1.89 &   1.67\\
  24000.0 &  2.07 &   1.85 &   1.75 &   1.60 &   1.75 &   1.53\\
  25000.0 &  1.90 &   1.66 &   1.57 &   1.42 &   1.57 &   1.37\\
\noalign{\smallskip}
\hline
\end{tabular}
\end{center}
\end{table}

\section{Elementary considerations for the accretion/diffusion
  scenario} The possibility of this scenario providing an explanation
of the observed metals in cool white dwarfs was studied extensively in
a series of three papers by Dupuis and collaborators
\citep{Dupuis.Fontaine.ea92, Dupuis.Fontaine.ea93,
  Dupuis.Fontaine.ea93*b}. In the first two of those papers, the
authors solved the complete problem of time-dependent accretion and
diffusion, by considering the spatial and temporal distribution of a
heavy element during the white dwarf cooling evolution. They also
showed that for {\em trace} elements in the outer convection zone --
which is always homogeneously mixed -- the computation can be replaced
by a much simpler local calculation at the bottom of the convection
zone. We use this approximation here to derive some basic consequences
of debris disk accretion onto white dwarfs, which are not always
appreciated in the current literature.

The basic equation for the mass abundance $X_\mathrm{cvz}(el)$ of heavy
element $el$ in the convection zone becomes \citep[see Eq.~1 in][with
slightly different notation]{Dupuis.Fontaine.ea92}
\begin{eqnarray} M_\mathrm{cvz}\,\frac{dX_\mathrm{cvz}(el,t)}{dt} 
&=& \dot{M}(el) - 4\pi r^2\, X_\mathrm{cvz}(el,t)\, \rho\, v(el)\nonumber \\
&=& \dot{M}(el) - \frac{X_\mathrm{cvz}(el,t)\,M_\mathrm{cvz}}{\tau_\mathrm{diff}}
\end{eqnarray}
where the first term on the right side is the accretion rate of that
element, and the second term is the rate of gravitational settling
(where we use a positive velocity to represent elements diffusing
downward). Assuming that $\tau, \dot{M}, v$ are constant, the solution
of this equation is
\begin{equation} 
X_\mathrm{cvz}(el,t) = X_\mathrm{cvz}(el,0)\,{\mathrm e}^{-t/\tau(el)} + 
 \frac{\tau(el)\, \dot{M}(el)}{M_\mathrm{cvz}}\,\left[ 1 - {\mathrm
     e}^{-t/\tau(el)} \right] 
\end{equation}
where the first term on the right side is the starting value of the
element abundance at the beginning of the accretion phase.

Some conclusions can be drawn immediately from this simple
equation. If the accretion rate is constant for some multiple of the
diffusion timescale ($\approx 5$ for practical purposes, given the
typical uncertainties in observed elemental abundances), the abundance
will approach an asymptotic (``steady state'') value (we omit the
index $\mathrm{cvz}$ from $X$ for simplicity) of
\begin{equation} X(el,\infty) = \frac{\tau(el)\, 
\dot{M}(el)}{M_\mathrm{cvz}}
\end{equation}
and for the ratio of elements 1 and 2, we obtain
\begin{equation} 
\frac{X(el1)}{X(el2)} = \frac{\tau(el1)\, \dot{M}(el1)}{\tau(el2)\, 
  \dot{M}(el2)} = \frac{\tau(el1)}{\tau(el2)}
\,\frac{X_\mathrm{acc}(el1)}{X_\mathrm{acc}(el2)}, 
\end{equation}
where in the last step, we replaced the accretion rate by the
element abundances in the accreted matter.  If the diffusion
times are less than a few years, we can reasonably assume that this
steady state has been reached. Since we can calculate the timescales
from the parameters of the star, we can thus determine the abundances
in the accreted matter from the observed abundances in the star. Even
more simply, since the timescales for the commonly observed elements
of Ca, Mg, and Fe are often within a factor of two, we can assume the observed
abundances (which are rarely more accurate than a factor of 2) to be a
first approximation to the accreted abundances.

However, all of these conclusions depend critically on the assumption
that the steady state is reached, and are therefore valid only in the
case of short diffusion timescales. It is instructive to consider by
a simple example, the consequences of being unable to assume
steady state  (i.e., in all cases where the timescales exceed
a few years).

For this exercise, we studied the abundances of two elements, where the
diffusion timescales differed by a factor of two. We started with zero
abundance and switched accretion on for 10 times the shorter
timescale. Then accretion is switched off again.

\begin{figure}
\includegraphics[width=8.8cm]{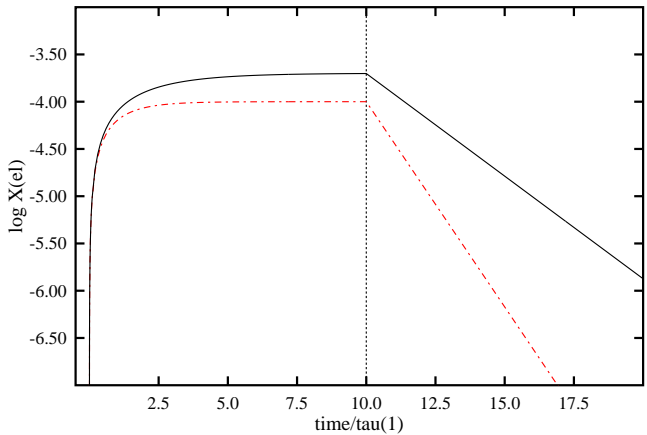}
\caption{Simple example for the accretion/diffusion scenario with two
  elements.
  \label{fig1}}
\end{figure}

We were able to distinguish three phases (see Fig.~\ref{fig1})
\begin{enumerate}
\item for $t \ll \tau_1,\tau_2$, expansion of the exponentials in the
  equation above shows that the diffusion timescales cancel; the
  element abundances increase linearly with time, with the slope given
  by the accretion rates (or the abundances in the accreted matter)
\item for $ t \approx \tau $, the abundances approach the steady state
  values, which are reached after $\approx 5 \tau$. In this phase, the
  abundances of the accreted material are modified by the ratio of the
  diffusion timescales, which is here equal to 2
\item when the accretion is switched off, the abundances decrease
  exponentially. Because of the difference in diffusion timescales,
  the ratio also increases exponentially and can reach very large
  values -- in this simple case $>100$ after 10 diffusion
  timescales. At this point, the element abundances are only 2-4 orders
  of magnitude below their maximum values, which would be observable
  in many cases.
\end{enumerate}

The third phase would also describe a different scenario, where 
accretion occurs at a high rate in a short time, such as  the
accretion of an entire asteroid at once. The abundances would reflect
exactly the accreted abundances during this time and start the
exponential decay from there. In such a case we would be unable able to
determine the composition of the accreted matter.

We note that the above example is not extreme. In particular,
when including light or very heavy elements in the comparison the
factor between timescales can be $\approx 4$. And we also do not need
to consider the extreme case of totally switching accretion on and
off. Even a change in the accretion rate, for example by a factor of
100 leads to an intermediate change in the abundance ratio of a factor
$>10$, before the asymptotic ratio of 2 is reached again.

The conclusion of this simple exercise is that it is only possible to
infer abundances of the accreted matter {\em if we can safely assume
  that the accretion rate has been constant for several times the
  diffusion timescales of all elements involved, and that therefore
  the steady state has been reached}. For diffusion timescales longer
than a few decades at most this can never be assumed. Equating the
observed atmospheric abundance ratios to those of the accreted
matter can in such cases be incorrect by orders of magnitude. 

\section{Specific examples}
We apply the general results to two specific
objects with circumstellar material and atmospheric metal traces,
which have been discussed in numerous recent studies: G\,29-38 and
GD\,362.

\subsection{G\,29-38}
Based on three independent determinations
\citep{Liebert.Bergeron.ea05, Koester.Provencal.ea97, Voss06}, we use
the atmospheric parameters $T_\mathrm{eff}$\ = 11\,690~K, $\log g$\ =
8.11. An atmosphere/envelope model was calculated with these data and
Table~\ref{tabg29} collects the diffusion timescales and observed
abundances from \citet{Koester.Provencal.ea97} and
\citet{Zuckerman.Koester.ea03}. Unfortunately, there is some confusion
about the Ca abundance in the literature -- even using only studies
with contributions from this author. The reason is that the Vienna
Atomic Line Database (VALD) contains Stark broadening data for the
\ion{Ca}{ii} H and K resonance lines, which differ by about a factor of
10 from most of the Kurucz lists. Depending on which data are used,
the abundance differs by about 0.5 dex. Using simple estimates for
Stark broadening indicates that the larger broadening constants are
more likely correct and thus the lower Ca abundance is preferred.

For the model, we used the version with the more efficient
convection, which provides a reasonable thermal timescale of about
100~s. Since the diffusion timescales are still smaller than one
year, the conclusions are not qualitatively changed compared to the
standard version, which was used in \citet{Koester.Wilken06}.
There is a significant caveat, however: as discussed in a previous section,
even though the thermal timescale agrees with the pulsation
properties, the possible overshooting at the bottom might increase the
diffusion timescales significantly. Another effect in the same
direction is the pulsation itself. During the pulsational decrease in
the effective temperature \citep[up to 500~K;][]{Montgomery05}, the
depth of the convection zone increases significantly, leading to a
larger mixed zone and an increase in the diffusion timescale by almost
a factor of 10 compared to the equilibrium model.

Metals were first detected in G\,29-38 in 1997
\citep{Koester.Provencal.ea97}. Von Hippel \& Thompson (2007) claimed
changes had occurred in the equivalent width of the \ion{Ca}{ii}
resonance lines during the past decade, which was questioned by
\citet{Debes.Lopez-Morales08}. We have seven high-quality
high-resolution spectra from the ESO VLT and the Keck telescope, which
show no significant variation between 1997 and 2000; the
\citet{von-Hippel.Thompson07} result may have been influenced by the
inhomogeneity of their data, which included time-resolved spectroscopy
of very low S/N.

\begin{table}[ht]
\caption{Diffusion data for the variable ZZ Ceti star G\,29-38. Column 2
  gives the time scale in years, column 3 the observed number
  abundances relative to Ca, column 4 the observed abundances by mass,
  also relative to Ca. Column 5 is the predicted abundance in the
  accreting material, {\em assuming  steady state between accretion and
  diffusion}. The final column shows the solar abundances on the same
  scale. All abundances are in decadal logarithms. 
\label{tabg29}}
\begin{center}
\begin{tabular}{rrrrrrr}
  \hline
  \noalign{\smallskip}
  el  & $\tau$ [yrs] & $\log \epsilon_\mathrm{atm}$ & 
$\log X_\mathrm{atm}$ & $ \log X_\mathrm{acc}$ & $\log X_\odot$\\
  \noalign{\smallskip}
  \hline
  \noalign{\smallskip}
 Ca    & 0.79 & 0.00  &  0.00 & 0.00 & 0.00 \\
 Mg    & 0.97 & 1.10  &  0.88 & 0.79 & 1.00 \\
 Fe    & 0.80 & 0.61  &  0.82 & 0.81 & 1.33 \\
\noalign{\smallskip}
\hline
\end{tabular}
\end{center}
\end{table}

In view of the short diffusion timescales, it seems thus reasonable to
assume that this object at the moment is experiencing steady state
accretion. Because of the similar timescales, the predicted
accretion abundances do not differ too much from the observed
values. Taking into account that there is a factor of 3 difference in the
Mg abundance between \citet{Koester.Provencal.ea97} and
\citet{Zuckerman.Koester.ea03} and a similar uncertainty in the Fe
abundance (high vs. low dispersion spectra, optical vs. UV spectra),
we can conclude that within the rather large uncertainties, the
abundances of Ca, Fe, and Mg in the accreted matter are compatible with
solar abundances, taken from the compilation in Astrophysical
Quantities \citep{Cox00}.

\subsection{GD\,362}
GD\,362 was discovered as a massive DA with metals (DAZ) by
\citet{Gianninas.Dufour.ea04}. An infrared excess
\citep{Kilic.von-Hippel.ea05} placed it into the new class of DA white
dwarfs with debris disks. However, an analysis of high S/N,
high-resolution spectra \citep[][Z07
  henceforth]{Zuckerman.Koester.ea07} showed it to be a normal mass,
but helium-rich object. We used the stellar parameters for an
atmosphere/envelope model, and inferred the diffusion parameters in
Table~\ref{tabgd362}. We found that the fractional mass in the convection zone is $
q = -6.58$, and, using the mass 0.73~M$_\odot$\ from Z07, that the total hydrogen
mass in the envelope is $3.51\times 10^{-9}$~M$_\odot$, and the total mass of
Si is $3.90\times 10^{21}$~g. The combined mass of the clearly identified
elements Mg, Al, Si, Ca, Fe, and Ni is $\approx 1.8\times 10^{22}$~g, in
excellent agreement with the estimates in Z07.

\begin{table}[ht]
\caption{Diffusion data for GD\,362. The meaning of the columns is the
  same as in Table~\ref{tabg29}, except that all element abundances
  are presented relative to Si.
\label{tabgd362}}
\begin{center}
\begin{tabular}{rrrrrrr}
  \hline
  \noalign{\smallskip}
  el  & $\tau$ [yrs] & $\log \epsilon_\mathrm{atm}$ & 
$\log X_\mathrm{atm}$ & $ \log X_\mathrm{acc}$ & $\log X_\odot$\\
  \noalign{\smallskip}
  \hline
  \noalign{\smallskip}
  C    & 1.28\,$10^5$ & $<$0.20 & $<$-0.17& $<$-0.38 &  0.64 \\
  N    & 1.22\,$10^5$ & $<$1.70 & $<$1.39 & $<$ 1.20 &  0.20 \\
  O    & 1.13\,$10^5$ & $<$0.70 & $<$0.45 & $<$ 0.30 &  1.14 \\
 Na    & 9.50\,$10^4$ &  -1.95  &   -2.04 &    -2.11 & -1.33 \\
 Mg    & 9.44\,$10^4$ &  -0.14  &   -0.21 &    -0.28 & -0.04 \\
 Al    & 8.57\,$10^4$ &  -0.56  &   -0.60 &    -0.63 & -1.10 \\
 Si    & 7.95\,$10^4$ &   0.00  &    0.00 &     0.00 &  0.00 \\
 Ca    & 6.07\,$10^4$ &  -0.40  &   -0.25 &    -0.13 & -1.04 \\
 Sc    & 5.83\,$10^4$ &  -4.35  &   -4.15 &    -4.02 & -4.24 \\
 Ti    & 5.45\,$10^4$ &  -2.11  &   -1.88 &    -1.72 & -2.33 \\
  V    & 5.20\,$10^4$ &  -2.90  &   -2.64 &    -2.46 & -3.29 \\
 Cr    & 5.05\,$10^4$ &  -1.57  &   -1.31 &    -1.11 & -1.61 \\
 Mn    & 4.77\,$10^4$ &  -1.63  &   -1.34 &    -1.12 & -1.87 \\
 Fe    & 4.67\,$10^4$ &   0.19  &    0.49 &     0.72 &  0.29 \\
 Co    & 4.43\,$10^4$ &  -2.66  &   -2.34 &    -2.09 & -2.31 \\
 Ni    & 4.45\,$10^4$ &  -1.23  &   -0.91 &    -0.66 & -0.98 \\
 Cu    & 4.25\,$10^4$ &  -3.34  &   -2.99 &    -2.72 & -2.98 \\
 Sr    & 3.14\,$10^4$ &  -4.58  &   -4.09 &    -3.69 & -4.15 \\
\noalign{\smallskip}
\hline
\end{tabular}
\end{center}
\end{table}

Given that the individual abundances vary over more than four orders
of magnitude, it is remarkable, how close the inferred accretion
abundances (column 5) are to the solar abundance ratios. The error for
Si was given as 0.3 dex in Z07, and for the other elements, it ranges
from 0.1 to 0.4 dex. The only elements with abundances that differ
significantly from the solar value are C and O, for which only upper
limits are known, and Ca and Na.  Since 12 of 14 elements have a ratio
of their abundance to that of Si higher than for solar values might
imply that the Si abundance in Z07 is slightly too low, producing a
systematic shift.

\begin{figure}
\includegraphics[width=8.8cm]{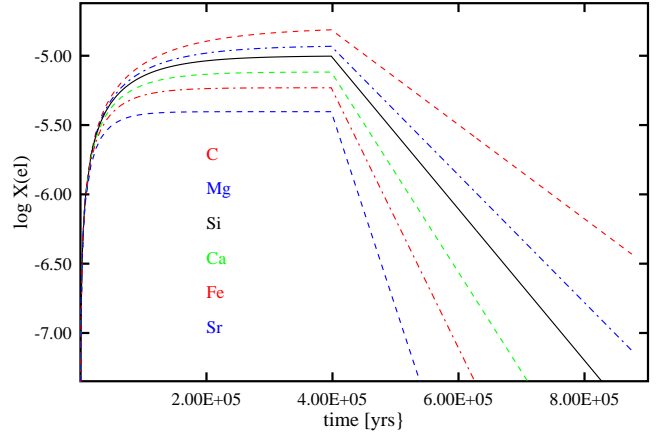}
\caption{Possible accretion/diffusion history for GD\,362. 
The element symbols are ordered in the same sequence as the corresponding
curve. 
\label{fig2}}
\end{figure}

The possible ``abundance history'' of GD\,362 is shown in
Fig.~\ref{fig2}, which is in a similar spirit as Fig.~\ref{fig1}. We
use the elements C, Mg, Si, Ca, Fe, and Sr as representatives for
different diffusion timescales and abundances.  Accretion is switched
on at time $t = 0 $, when no heavy elements are present in the
atmosphere; the rate -- the same for all elements -- is calculated to
ensure a steady state abundance of silicon of $\log X(\mathrm{Si})=
-5$, the observed value. The silicon rate necessary for this is
$2.4\times 10^{17}$\,g per year. A different rate would shift all
curves vertically by the same amount. In reality, the elements would
of course be accreted with different abundances. In our choice of
presentation, the difference between the curves shows directly the
relative change between the accreted and atmospheric abundance ratios.
When all the curves are approximately in agreement, the observed
abundance ratios are identical to the accreted ratios. In this
specific case, this is true for times $ t \la 3\times 10^4$\,yrs. After that
time, the steady state is approached, and the ratio differs from the
accreted ones by the ratios of the diffusion timescales, which are factors of
2-4 in this case.

After the accretion is switched off, the picture changes
significantly. About $4\times 10^5$\,yrs after the end of accretion, the
ratios can differ from the accreted abundance ratios by factors of
up to 200. This decline is fast on astronomical timescales, but slow
enough to be invisible over a few decades; we thus do not have any
direct indication of the phase we are currently observing. It
should also be noted -- as in the simple example above -- that
a change in the accretion rate of a factor of 10-100 would force the
star into a phase of exponential decline, until a new steady state
with lower abundances was reached.

In their study of accretion from the interstellar matter,
\cite{Dupuis.Fontaine.ea93} assumed schematically a duration of the
accretion phase within a dense ISM cloud of $10^6$\,yrs. This would
correspond to the time required to complete all of the process indicated in 
Fig.\ref{fig2}, and all phases would be possible for the current
situation. On the other hand, in the scenario of accretion from the
debris disk formed by an asteroid (see e.g., Z07 for an extensive
discussion), the total accreted matter of the identified elements
alone would be $\approx 7.5\times 10^{23}$\,g after $3\times 10^5$\,yrs, close to
the mass of Ceres, the most massive asteroid in our system. This
agrees with estimates of the lifetime of the debris disks, which are of the
order of $1.5\times 10^5$\,yrs \citep{Jura08, Kilic.Farihi.ea08}.

It is thus  highly likely that we are witnessing the early phase of
accretion, and that it will be possible to identify the composition of the
accreted matter. This view is further supported by the visibility of
rare elements such as strontium, which have short diffusion timescales. Within
$\approx 10^5$\,yrs of the end of the accretion phase, the Sr
abundance would decrease by an order of magnitude, becoming
undetectable. 

We also note that the aforementioned deviations from solar abundances
for C, N, Na, and Ca cannot be explained by assuming a post-accretion
phase. Deviations would be in the opposite direction of expectations:
C, O, and Na are underabundant, but should increase in this phase, and
vice versa for Ca.

An element of particular importance for the accretion/diffusion
scenario is hydrogen, because as the lightest element, it will always
remain in the convection zone and not diffuse downward. Assuming that
hydrogen would be accreted with solar ratio with respect to silicon, the
current hydrogen mass present -- $6.98\times 10^{24}$\,g -- would be
reached in 16\,580\,yrs. We either are observing GD\,362 in the very early
phase of accretion, or the accreted matter is significantly
hydrogen-depleted.

\begin{figure}
\includegraphics[width=8.8cm]{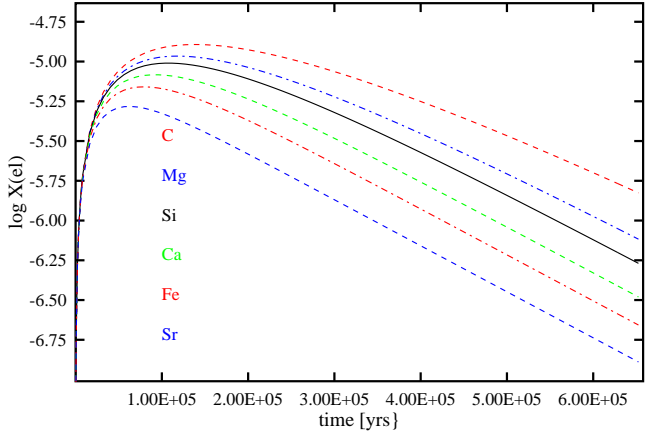}
\caption{Alternative history for GD\,362, assuming an exponentially
  declining accretion rate. Otherwise the meaning of the curves is the
  same as in Fig.~\ref{fig2}.
  \label{fig3}}
\end{figure}

\citet{Jura.Muno.ea09} considered an exponentially decreasing accretion
rate, which would be appropriate if the dust disk is not replenished
from a reservoir farther out in the system. In such a case, the
diffusion equation can also be solved analytically to have the solution
(in our notation)
\begin{eqnarray}
X_\mathrm{cvz}(el,t)& = & X_\mathrm{cvz}(el,0)\,{\mathrm
  e}^{-t/\tau(el)} \nonumber \\
& + & \frac{\tau(el)\,   \dot{M}_0(el)}{M_\mathrm{cvz}}\,
\frac{\tau_\mathrm{acc}}{\tau_\mathrm{acc}-\tau(el)}  \,
\left[{\mathrm e}^{-t/\tau_\mathrm{acc}}  - {\mathrm e}^{-t/\tau(el)} \right] 
\end{eqnarray}
with a timescale for the accretion rate of $\tau_\mathrm{acc}$ and an
initial rate $\dot{M}_0(el)$. We calculated the ``abundance
history'' for GD\,362 with the smaller of the two times considered in
\citet{Jura.Muno.ea09}, $\tau_\mathrm{acc} = 1.5\times 10^5$\,yrs, and show
the results in Fig.~\ref{fig3}.

In this case, the abundances never reach an asymptotic state, but
instead begin to increase, reach  a maximum, and then decline. The
maxima are reached at different times for the elements, depending on
their diffusion timescales. The initial accretion rate for Si (and
all other elements) had to be increased by a factor of two from
the case of constant rate, because otherwise the Si abundance would
never reach the observed value. 

In this example, the accretion timescale is longer than all diffusion
timescales. As time increases, the later decline is governed by the
accretion timescale, which is the same for all elements. The abundance
ratios therefore do not show the large deviations from the accreted
values as in the case, where accretion is switched off. Instead of the
ratios of the diffusion timescales that govern the relative changes in
the ``steady state'', we obtain here the ratio
\begin{equation}
 \frac{X(el1,\infty)}{X(el2,\infty)} =
 \frac{\tau(el1)\, X_\mathrm{acc}(el1)}{\tau(el2)\, X_\mathrm{acc}(el2)}
 \,\frac{\tau_\mathrm{acc} - \tau(el2)}{\tau_\mathrm{acc} - \tau(el1)} ,
\end{equation}
which for very large $\tau_\mathrm{acc}$ relative to $\tau(el)$
recovers the solution for the steady state.

If the accretion timescale is shorter than the diffusion timescale
of some elements, then at later times for these elements the
conditions approach an exponential decline after the
switching off of accretion, and there is the possibility of large deviations from
the original abundance. This might in the future be helpful in providing 
constraints on the lifetimes of the disks.

\section{Conclusions}
We have extended the calculation of diffusion timescales to many more
elements than currently available in the literature, to provide the
data necessary to interpret the observed abundances in the increasing
number of white dwarfs with traces of metals in their atmospheres and
an infrared excess, indicative of the presence of a dust disk. With a
simple example and the application to two of the most extensively
observed objects (G\,29-38 and GD\,362), we have demonstrated that the
observed abundances are in general not identical to the accreted
abundances. If, however, we have good arguments that the steady state
between accretion and diffusion is reached, the differences are
typically only a factor of 2-4, in many cases within the errors of the
determination. This is still true, if the accretion rate changes only
slowly, on timescales longer than all diffusion timescales
involved. With knowledge of the timescales, we can improve the
comparison by calculating the accreted abundances from the observed
ones. With further observations of the highest quality, it should thus
be possible to resolve the question about the origin of the accreted
matter.

\begin{acknowledgements}
I gratefully acknowledge very helpful discussions with Mike Jura and
the opportunity to read a manuscript in advance of publication. In
particular the consideration of exponentially decreasing accretion
rates was stimulated by his ideas.
\end{acknowledgements}

\end{document}